\documentstyle[epsfig]{aipproc}

\begin{document}

\title{Multidimensional Mesh Approaches to Calculations of Atoms
and Diatomic Molecules in Strong Electric Fields}

\author{Mikhail V. Ivanov}
\address{Theoretische Chemie, Physikalisch--Chemisches Institut,\\
Universit\"at Heidelberg, INF 229, D-69120 Heidelberg,\\
Federal Republic of Germany\\
e-mail: Mikhail.Ivanov@tc.pci.uni-heidelberg.de\\
Institute of Precambrian Geology and Geochronology,\\
Russian Academy of Sciences,\\
Nab. Makarova 2, St. Petersburg 199034, Russia\\
e-mail: MIvanov@MI1596.spb.edu
}

\maketitle

\begin{abstract}
Fully numerical mesh solutions of 2D and 3D quantum equations of
Schr\"odinger and Hartree-Fock type allow us to work with
wavefunctions which possess a very flexible geometry.
This flexibility is especially important
for calculations of atoms and molecules in strong external fields
where neither the external
field nor the internal interactions can be considered as a perturbation.
In the framework of this method
we present various approaches to calculations of quasi-steady states
of these systems in strong electric fields.
These approaches are aimed at obtaining precise complex wavefunctions
and corresponding eigenvalues in the form $E = E_0 - i\Gamma/2$,
where $E_0$ is the real part of the energy and the value $\Gamma/2$
determines the lifetime of the state in relation to escape
of electrons from the system.
The applications for single-electron
systems include correct solutions of the Schr\"odinger equation
for the ${\rm H_2^+}$ ion (energies
and decay rates) and the hydrogen atom in strong
parallel electric and magnetic fields.
Some results for the helium atom in strong electric fields 
are announced.
\end{abstract}

\section{Introduction}

During the latter decade the interest to theoretical studies 
of atoms and molecules
in strong external fields was strongly 
motivated by experiments with intense laser
beams (electromagnetic fields with dominating electric component)
and astronomical observations of white dwarfs and neutron stars
(magnetic fields).
The experimental availability of extremely strong electric fields in laser
beams makes the theoretical study of various
atomic and molecular species under such conditions
very desirable.
The properties of atomic and molecular systems in
strong fields undergo dramatic changes in comparison with
the field-free case.
These changes are associated with the strong distortions of
the spatial distributions of the electronic density and correspondingly
the geometry of the electronic wavefunctions.
This complex geometry is difficult for its description by means
of traditional sets of basis functions and requires
more flexible approaches which can, in particular,
be provided by multi-dimensional mesh
finite-difference methods.

Many 
results of the experiments with intense laser beams can be
considered from the point of view of the behaviour of atoms and
molecules in intense static electric fields, especially when the
frequency of the radiation is low. (The low-frequency behaviour and the
limits of this region are analysed by
\cite{Keldysh64}).

The most advanced theoretical
studies of effects in strong electric fields were 
traditionally concentrated on a
hydrogen atom. Other atomic and molecular systems in strong
electric fields are much less studied. Many sophisticated
theoretical methods developed for the hydrogen atom cannot be
simply applied for other atoms and molecules. This circumstance is
an argument for development of more universal theoretical and
computational methods for atoms and molecules in strong electric
fields.

Quasi-steady states of hydrogen
atoms in strong electric fields were studied precisely in
many theoretical works (see 
\cite{BenGre80,Franceschini,Froelich76,Kolosov87,Kolosov89,NicThem92,NicGot92,AliHopf94,Rao94,SilvermaN88}
and references
therein). Some of these works are based on separation of variables in
parabolic coordinates in the Schr\"odinger equation for the
hydrogen atom. This separation of spatial variables for
atomic and molecular systems in external electric fields is
possible only for one-electron atoms with the Coulomb
electron-nucleus interaction. Non-hydrogenic systems or
systems with non-Coulomb interaction do not allow this
separation of variables. 
Their
theoretical studies require a developed technique of solving
the Schr\"odinger and similar equations with non-separable variables. 
One
of the possible ways for solution of this problem consists
in the application of mesh computational methods for solving 
these equations.
The mathematical problem consists here in solution of partial
differential equations for systems with discrete
quasi-steady states lying on the background of the
continuous spectrum.

In the first part of this work we present applications of our 
mesh method for solving Schr\"odinger
equations with non-separable variables for
quasi-steady states. The method is based on the
technique of a mesh numerical solution of
Schr\"odinger and Hartree-Fock equations with
non-separable variables for steady states 
\cite{Ivanov82,ZhVychMat,Ivanov88,Ivanov91,Ivanov94,Ivanov98}. 
The most of the applications for the discrete states 
was focused on atoms in strong magnetic fields 
\cite{Ivanov88,Ivanov91,Ivanov94,Ivanov98,IvaSchm98,IvaSchm99,IvaSchm2000}. 
In this paper we present computational approaches which 
we have developed for quasi-steady states in external electric fields 
\cite{Ivanov94a,Ivanov98a,Ivanov2001} 
and apply them to several single electron systems. 
At the end of the paper we announce some of results 
of our current work on the helium atom in strong electric fields.

\section{Formulation of the problem and the 2D mesh computational method 
for stationary states}

We carry out our mesh solution in the cylindrical coordinate system
$(\rho,\phi,z)$ with the $z$-axis oriented along the electric field.
After separation of the $\phi$ coordinate the Hamiltonians of 
the single-electron problems 
considered below take the form 
\begin{eqnarray}
H=-\frac 12\left(\frac{\partial ^2}{\partial \rho ^2}+
\frac 1\rho \frac \partial
{\partial \rho }+\frac{\partial ^2}{\partial z^2}-
\frac{m^2}{\rho ^2}\right)
+s_z\gamma+\frac{m}{2}\gamma+\frac{\gamma^2}{8}\rho^2
-\frac 1r-Fz
\label{HFHam}
\end{eqnarray}
-- the hydrogen atom in parallel electric and magnetic fields and
\begin{eqnarray}
H=
-\frac 12\left({\frac{\partial ^2}{\partial \rho ^2}+
\frac {1}{\rho} \frac \partial{\partial \rho}+
\frac{\partial ^2}{\partial z^2}-\frac{m^2}{\rho^2}}\right)
+s_z\gamma+\frac{m}{2}\gamma+\frac{\gamma^2}{8}\rho^2  \nonumber
\\
-
\frac {1} {\left[\left(z+R/2\right)^2 +\rho^2\right]^{1/2}}
-\frac {1} {\left[\left(z-R/2\right)^2
+\rho^2\right]^{1/2}}-Fz\label{HHHP}
\end{eqnarray}
-- the molecular ion ${\rm H_2^+}$ in electric and magnetic
fields parallel to the molecular axis.
Here and in the following
the atomic units $m_e=\hbar=e=1$ will be used, including the
magnetic field strength $\gamma$ measured in units
$B_0=\hbar c/ea_0^2=2.3505 {\cdot} 10^5$T,
$\gamma =B/B_0$.
$F$ is the electric
field strength multiplied by the charge of the electron. 
$F=1$ corresponds to $51.422{\rm V/\AA}$. 
The value $m$ is the magnetic quantum number
and $s_{z}=\pm {1\over2}$ is the spin $z$-projection.
$R$ is the internuclear distance for ${\rm H^+_2}$ molecule.
The Hamiltonian (\ref{HHHP}) does not include the internuclear repulsion
and its eigenvalues are the electronic energies $E_e$ which are connected
with the total energy of  ${\rm H^+_2}$ by the formula $E=E_e+1/R$.

Our two-dimensional mesh computational method which we apply for 
obtaining eigenvalues of the Hamiltonians (\ref{HFHam}) and (\ref{HHHP}) 
in the form sufficient for case of stationary states (i.e $F=0$) 
of single-electron systems is described 
in refs.\cite{ZhVychMat,Ivanov88}. 
For these systems highly precisely solutions can be obtained by 
solving Schr\"odinger equations in finite spatial domains $\Omega$
(with simple boundary conditions $\left. \psi_{_{}} \right|_{\partial\Omega}=0$
or $\left. \partial \psi_{_{}}/\partial n \right|_{\partial\Omega}=0$)
with negligible errors for
domains of reasonable dimensions (see analytical estimations of errors 
in ref.\cite{Ivanov88} and results of numerical experiments 
in ref.\cite{ZhVychMat}).
When employing the Richardson's extrapolation of the energy values  
to the infinitely small mesh step $h\rightarrow 0$ very precise  
results can be obtained on uniform meshes with relatively 
small numbers of nodes \cite{ZhVychMat,Ivanov88}.
Two important problems have been solved 
for the following development of the multi-dimensional Hartree-Fock 
method for many-electron systems: 1. An enhancement of the precision 
of single-electron wavefunctions due to their more complicated geometry 
in comparison with the one-electron case, and 
2. Obtaining correct mesh representations of Coulomb and exchange potentials. 
The first of these problems can be solved more or less simply 
by means of non-uniform meshes with the distribution of nodes 
concentrated near the nuclei. 
The second and most complicated problem was initially 
solved in our first works on the helium atom in magnetic fields
\cite{Ivanov91,Ivanov94} by means
of a direct summation over the mesh nodes.
But this direct method is very expensive with respect to 
the computing time and due to this reason we obtained in the following
works \cite{Ivanov98,IvaSchm98,IvaSchm99,IvaSchm2000}
these potentials as solutions
of the corresponding Poisson equations.
The problem of the boundary conditions for a Poisson equation
as well as
the problem of simultaneously solving
Poisson equations on the same meshes with
Schr\"odinger-like equations for the wave functions $\psi_\mu(z,\rho)$
has been discussed in ref.\cite{Ivanov94}.

The simultaneous solution of the Poisson equations for the
Coulomb and exchange potentials and the 
Schr\"odinger-like equations for the wave functions $\psi_\mu(z,\rho)$
is a complicated problem due to a different asymptotic behaviour of
the wavefunctions and potentials.
The wavefunctions of the bound electrons decrease exponentially
as $r\rightarrow \infty$ ($r$ is the distance from the origin).
This simplifies the problem of the solution of 
equations for wavefunctions in the infinite space 
because it is possible either to solve
these equations in a finite domain $\Omega$
(as described above) or otherwise
to solve these equations in the infinite
space on meshes with exponentially growing distances
between nodes as $r\rightarrow \infty$.
On the contrary, solutions of Poisson equations for non-zero sums of charges
decrease as $1/r$ as $r\rightarrow \infty$.
In result, every spatial restriction of the domain $\Omega$
introduces a significant error into the final solution.
In our approach we address the above problems
by using special forms of non-uniform meshes \cite{Ivanov98}.
Solutions to the Poisson equation on separate meshes contain
some errors $\delta_P$ associated with an inaccurate description of the
potential far from the nucleus.
However, due to the special form of the function $\delta_P(\tilde{h})$
for these meshes
(where $\tilde{h}$ is a formal mesh step)
the errors do not show up in the final results for the energy
and other physical quantities, which we obtain by means of the Richardson
extrapolation procedure (polynomial extrapolation to $h=0$
\cite{ZhVychMat,Ivanov88}).
The main requirement for these meshes is 
a polynomial increase of the actual mesh step $h(r)$ when $r\rightarrow \infty$.
Moreover, this behaviour can be only linear one,
i.e. $h^{-1}(r)=O(1/r)$ as $r\rightarrow \infty$.
The error of the mesh solution in this case has the form
of a polynomial of the formal step of the mesh $\tilde{h}=1/N$,
where $N$ is the number of nodes along one of the coordinates.
In practical calculations these meshes are introduced by means
of an orthogonal coordinate transformation from
the physical coordinates $x_{\rm p}$
to the mathematical ones $x_{\rm m}$ made separately
for $\rho$ and $z$.
The numerical solution is, in fact, carried out
on uniform meshes in the mathematical coordinates $x_{\rm m}$.
The characteristic feature of these meshes consists of rapidly increasing
coordinates of several outermost nodes when increasing the total number
of nodes and decreasing the actual mesh step in the vicinity of the origin.

The methods described above for the two-dimensional case can be applied 
also to three-dimensional Schr\"odinger and Hartree-Fock problems 
\cite{Ivanov85,Ivanov97}. 

\section{Quasistationary states}

\begin{figure}[b!]
\centerline{\epsfig{file=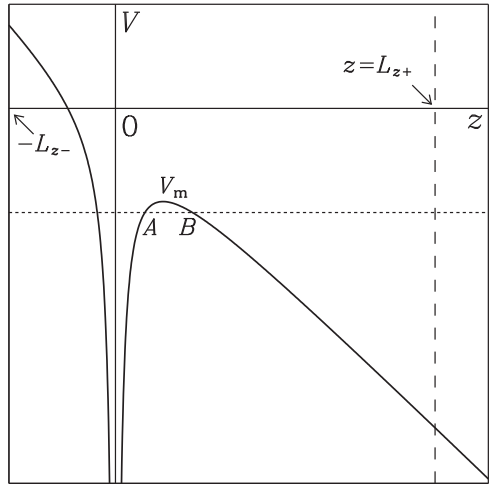,width=7cm}}
\vspace{10pt}
\caption{The potential energy for the hydrogen atom $V(\rho=0,z)$
in the external uniform electric field.}
\label{foobar:1r-fz}
\end{figure}

The most common approach to the mathematical description 
of the quasistationary states of quantum systems 
consists in employment of the complex form of the energy eigenvalues
\begin{eqnarray}
E=E_0-i\Gamma/2 \nonumber
\end{eqnarray}
where the real part of the energy
$E_0$ is the centre of the band corresponding to a quasistationary
state and the imaginary part
$\Gamma/2$ is the half-width of the band which determines the lifetime of
the state.
In this approach one may consider quasistationary states of quantum systems similarly
to the stationary ones. 
For systems (\ref{HFHam}) and (\ref{HHHP}) 
at $F>0$ the electron can leave the system in the direction of
positive $z$ and the behaviour of the wavefunction on this
semi-axis determines the main features of the behaviour in the
external electric field $F$.
The mathematical problem consists in the solution of the
Schr\"odinger equation
\begin{eqnarray}
H\psi =E\psi \label{eqnum}
\end{eqnarray}
at $F>0$ for resonance states which are inheritors of some
discrete (at $F=0$) states.
The wavefunctions of these states must describe the process of
separation of the electron from the system.
These wavefunctions can be distinguished either by the explicit
establishing the boundary condition of the outgoing wave or by
means of a complex coordinate transformation.
The latter transforms the
oscillating outgoing wave into an exponentially decreasing
wavefunction for which the simple Dirichlet boundary condition
$\psi\rightarrow 0$ as $z\rightarrow +\infty$ can be established.

From the mathematical point of view the problem under consideration consists
in obtaining solutions of the
single-particle Schr\"odinger equation for this electron with the correct
asymptotic behaviour of the wavefunction as an outgoing
wave.
Currently we have three different possibilities for obtaining solutions 
with the outgoing wave 
asymptotics realised in our computational program:

{\it 1. Complex boundary condition method.}
This method is described in detail in ref. \cite{Ivanov94,Ivanov98a}.
The method is based on the idea that the single-electron Schr\"odinger equation
for a finite system can be solved with the arbitrary precision
in a finite area both for stationary and for quasi-stationary eigenstates.
The case of stationary states is considered in \cite{ZhVychMat,Ivanov88}.
We discuss the approach for the quasi-stationary states following ref.\cite{Ivanov98a}.
Figure~\ref{foobar:1r-fz} presents the potential curve for the simplest
Hamiltonian (\ref{HFHam}). 
Analogously to \cite{Ivanov88,Ivanov83VLGU}
the calculations can be carried out in an area
$\Omega $ which is finite along the direction $z$.
For the coordinate $z$ we have used
uniform meshes.
The boundary of the area $z=-L_{z-}$ for
$z<0$ ($F\geq 0$) (Figure~\ref{foobar:1r-fz}) is determined from the
condition of small values of the wavefunction on the
boundary and, therefore, small perturbations introduced by
the corresponding boundary condition \cite{Ivanov88}.
The values of the wavefunction on the opposite boundary of
the area ($z=L_{z+}$) cannot be excluded from the
consideration. 
We consider non-stationary states 
corresponding to the 
process when an electron leaves the system in the direction
$z\rightarrow +\infty $. 
Thus, an outgoing wave boundary condition is to be
established on $z=L_{z+}$. 
The form of this boundary
condition can be derived from the asymptotic behaviour of the
wavefunction for $z\rightarrow +\infty $. 
In this limit the asymptotics of the real system can be replaced
by the asymptotics of the wavefunction of a free electron in the uniform
electric field. Solutions of the Schr\"odinger equation
for this system can be written as
\begin{equation}
\psi (z)=AM(\xi ) e ^{-i \Theta(\xi )}
{\rm , \ \ \ \ \ \ \ \ }
\xi =\left(z+\frac{E}{F}\right)(2F)^{1/3}
\label{Airyxi}
\end{equation}
where $M(\xi)$ and $\Theta(\xi)$ are the modulus and phase of the Airy
function, $A$ is a constant \cite{AbramSteg}. 
The
asymptotics of $\psi$ for $z\rightarrow +\infty $ can be obtained from
eqs. (10.4.78) and (10.4.79) of \cite{AbramSteg} in the
form
\begin{equation}
\psi(z)=\frac{A}{\sqrt{\pi}}\xi^{-1/4}
\exp\left(-i\frac\pi4+i\frac23\xi^{3/2}\right)+O\left(\xi^{-13/4}\right)
\label{Airyas}
\end{equation}
Taking into account that 
$d\xi/dz=(2F)^{1/3}$ 
and
$\xi^{1/2}(2F)^{1/3}=[2(E+Fz)]^{1/2}$ 
we have from (\ref{Airyas})
\begin{eqnarray}
\frac{d\psi}{dz}=\left( i k - \frac F{2k^2}
\right)\psi+\frac{d}{d\xi}O(\xi^{-13/4})\label{dpsidz}
\end{eqnarray}
where $k=[2(E+Fz)]^{1/2}$ is the wavenumber. Equation (\ref{dpsidz})
allows us to establish the following outgoing wave boundary condition
on the upper (in the $z$ direction) edge of the region
\begin{eqnarray}
\left. \frac{\partial \psi }{\partial z}+\left( \frac F{2k^2}-%
      i k\right) \psi\right| _{z=L_{z+}}=0\label{FeOutgo}
\end{eqnarray}
This approximate boundary condition is derived from the asymptotics of the
wavefunction of a free electron in a uniform electric
field and in the limit
$L_{z+}\rightarrow+\infty$ can be considered as exact one.
On the other hand, our numerical experiments show, that 
errors caused by the approximate nature of this asymptotics 
are important only for very short regions $\Omega$. 
Solving the Schr\"odinger equation with
the boundary condition (\ref{FeOutgo}) established on a reasonable
distances $L_{z+}$ from the origin of the system one obtains precise 
complex eigenvalues of the energy and corresponding wavefunctions.

This straightforward approach enables obtaining precise
results both for atoms and molecules from weak to moderate
strong fields (for instance for the ground state of the
hydrogen atom up to $F=0.20-0.25$~a.u.).
For stronger fields the precision of this method is limited 
by the precision of the mesh representation of the boundary 
condition (\ref{FeOutgo}).

\begin{figure}[b!] 
\centerline{\epsfig{file=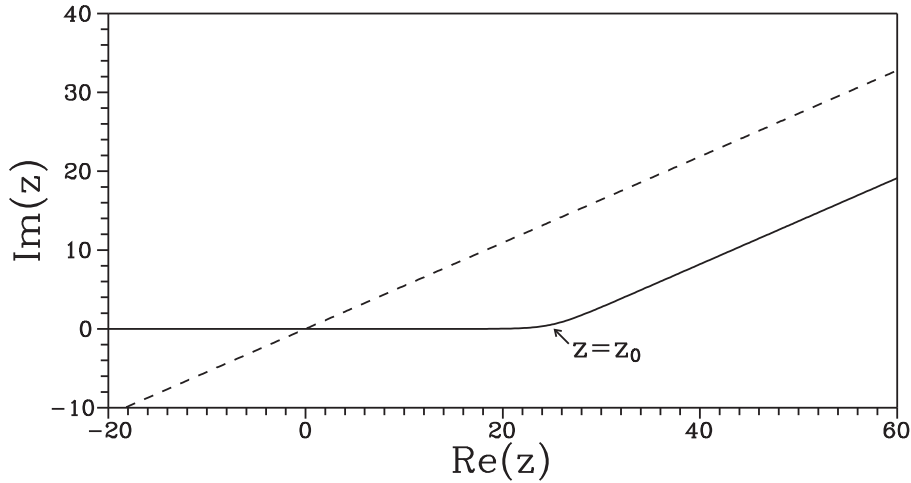,width=12cm}}
\vspace{10pt}
\caption{The integration paths for the complex coordinate $z$.
Dashed line is the path at the traditional rotation
$ z \rightarrow z e^{i \Omega} $.
Solid line is the path corresponded to the exterior curved path coordinate
transformation.}
\label{foobar:imre}
\end{figure}

{\it 2. Classical complex rotation of the coordinate $z$ in the
form $z \rightarrow ze^{i\Theta}$.}
Opposite to the boundary condition approach, 
this method does not require establishing boundary conditions 
dependent on the energy and can be used with the zero boundary conditions 
for the wavefunction. 
This simplification is especially important for applications 
to multi-electron systems.
In this approach we have obtained precise results for atomic
systems in strong fields from the lower bound of the over-barrier regime
up to super-strong fields corresponding to regime $|{\rm Re}E|<<|{\rm Im}E|$
\cite{Ivanov2001}.
For atoms in weak fields the applicability of the method 
is limited by numerical errors in the imaginary part of the energy. 
On the other hand, this method cannot be immediately applied
to molecular systems in our direct mesh approach \cite{Ivanov2001,Simon79}.

{\it 3. Exterior complex transformation of the coordinate $z$.}
The exterior complex scaling \cite{Simon79} combines many 
advantages of the boundary condition and complex rotation 
method. 
In its initial form \cite{Simon79} it consists in the complex 
rotation of a coordinate e.g. $r$ around a point $r_0$ 
lying in the exterior part of the system, i.e.
\begin{equation}
r\rightarrow r_0+(r-r_0)e^{i\Theta}{\rm \ \ \ for \ \ \ } r\geq r_0
\label{Simonform}
\end{equation}
and 
leaves intact the Hamiltonian in the internal part of the system. 
The latter circumstance allows us to employ this transformation 
both for atoms and molecules in arbitrary electric fields.  
In addition this transformation does not contain energy-dependent 
boundary conditions as well as the classical complex rotation. 
On the other hand, the exterior complex scaling being introduced 
into our numerical scheme
in the form (\ref{Simonform}) leads to the same numerical 
problems at very strong fields as well as the boundary condition method 
due to the nonanalytic behaviour of this transformation 
at the point $r_0$ (or $z_0$ in our case).

In our numerical approach we solve the latter problem by means 
of a  transformation of the real coordinate $z$
into a smooth curved path in the complex plane $z$ 
(Figure \ref{foobar:imre}, see details in \cite{Ivanov2001}). 
This transformation leaves intact the Hamiltonian in the
internal part of the system, but supplies the complex rotation of
$z$ (and the possibility to use the zero asymptotic boundary conditions
for the wavefunction) in the external part of the system 
without any loss of analycity.
The transformation can be applied both for atoms and molecules and
provides precise results for fields from weak up to super-strong 
including 
the regime $|{\rm Re}E|<<|{\rm Im}E|$
\cite{Ivanov2001}.

The three methods presented above have different but overlapping regions
of their most effective application.
This allows their combined using for the reciprocal control
in applications.
The numerical results obtained by all three methods coincide
in the limits of their applicability 
and are in agreement with numerous published data
on the hydrogen atom in electric fields (see e.g.
\cite{Kolosov87,Kolosov89,NicThem92,NicGot92}).

\section{Selected physical results}

\begin{figure}[b!]
\centerline{\epsfig{file=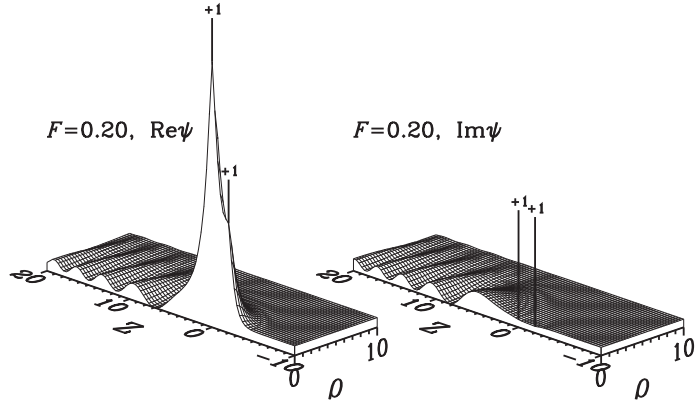,width=12cm}}
\vspace{10pt}
\caption{Real and imaginary parts
of the wavefunction of the $\rm H_{2}^+$
molecule in a longitudinal electric field $F$ (a.u.).}
\label{foobar:HHwave}
\end{figure}

\begin{figure}[b!] 
\centerline{\epsfig{file=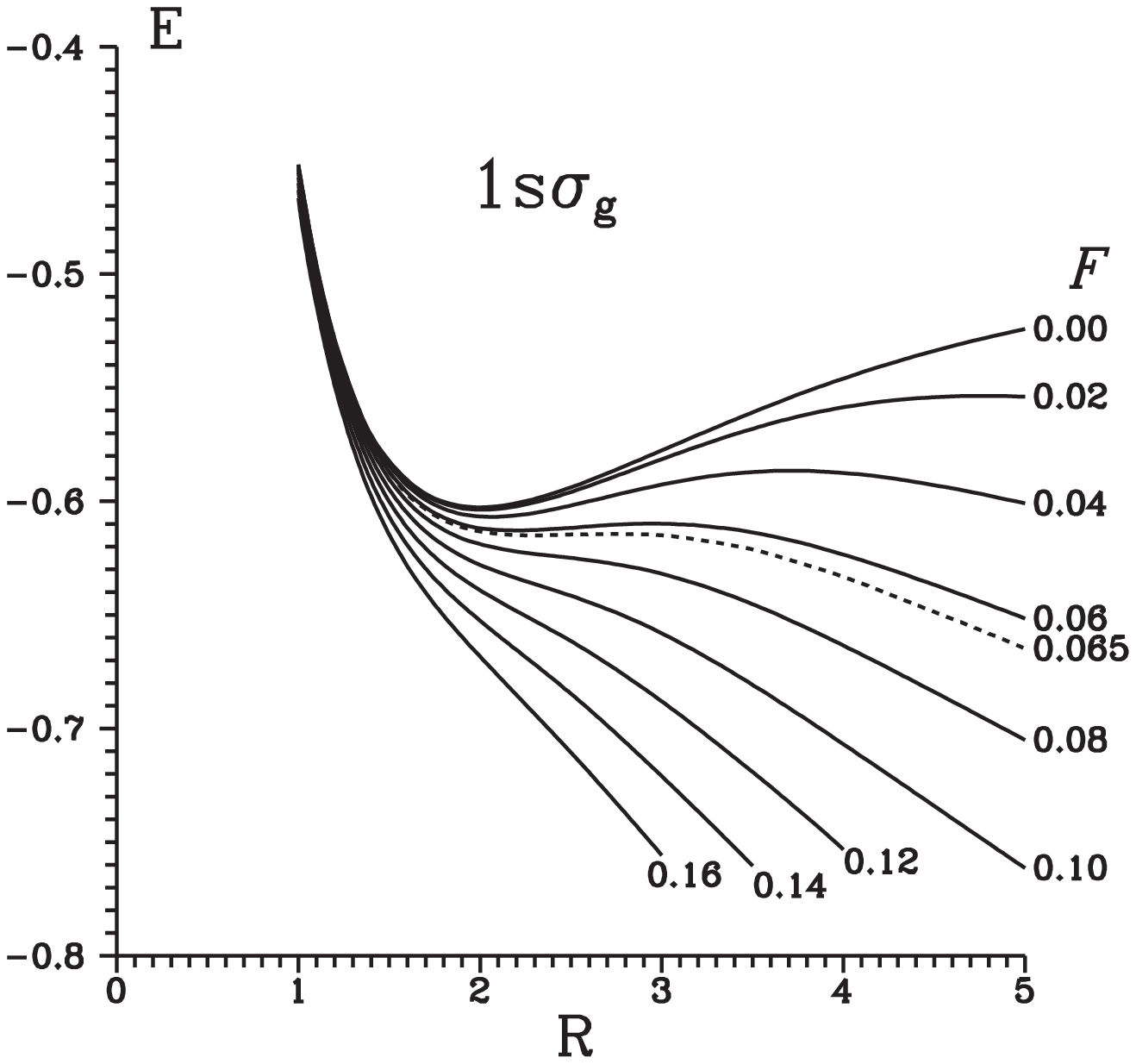,width=7cm}\hspace*{0.8cm}\epsfig{file=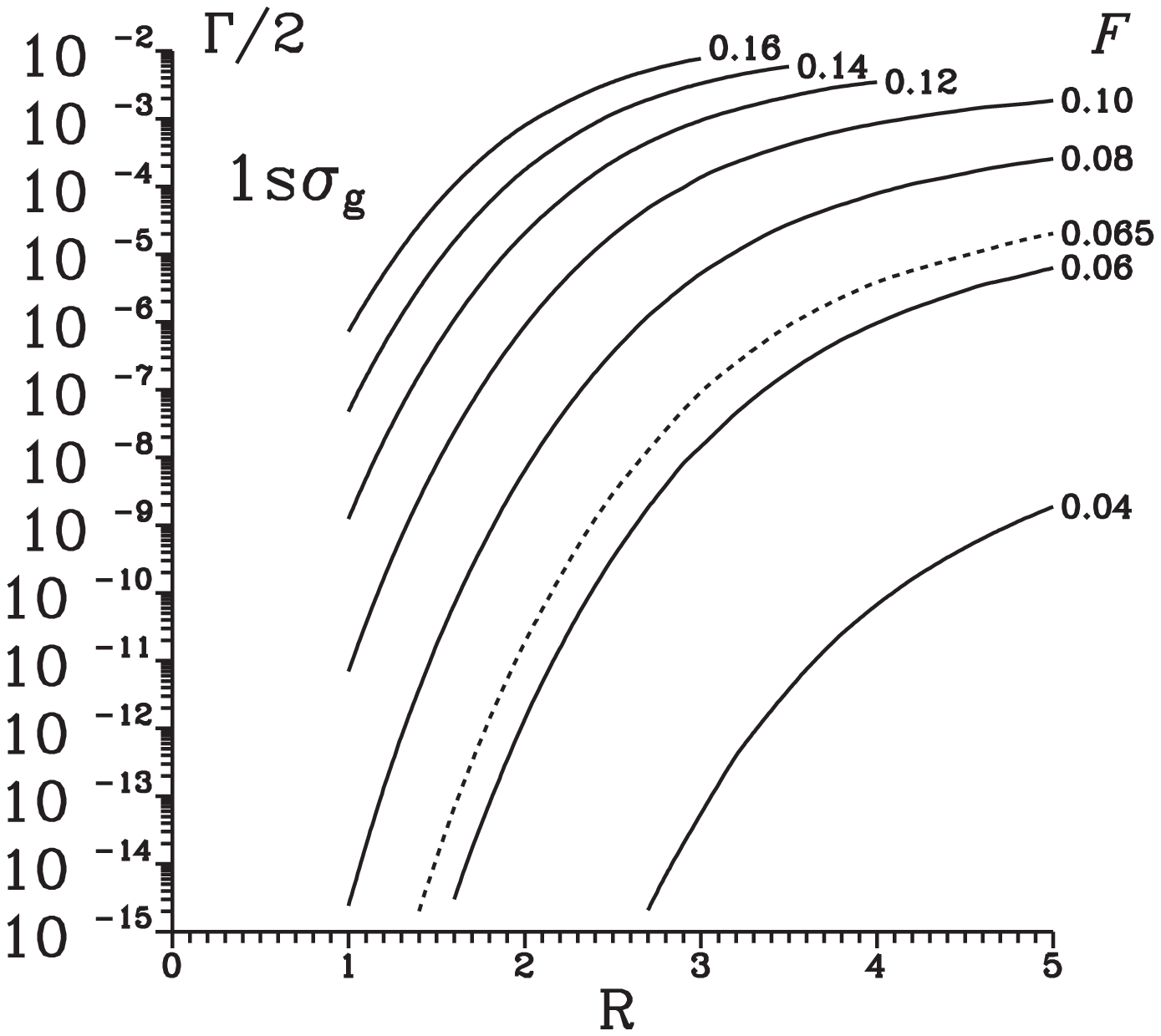,width=7cm}}
\vspace{10pt}
\caption{Left -- Potential curves for the ground state of the $\rm H_{2}^+$
molecule in the longitudinal electric field $F$ (a.u.).
Right -- Half-widths of the ground state energy level of the molecule
$\rm H_{2}^+$ in the longitudinal electric field $F$ as a function of the
internuclear distance $R$ (a.u.).}
\label{foobar:HHPEFGRA}
\end{figure}

Using the boundary condition approach enables one 
to obtain both the values of the energy and other 
observables and the wavefunctions in their mesh representation. 
An example of such a wavefunction of the ground state of the 
hydrogen molecular ion is presented in Figure~\ref{foobar:HHwave}.
The electron energy $E_{\rm e}$ obtained as
a solution of (\ref{eqnum}), (\ref{HHHP}) allows one to determine potential
curves $E(R)$ for the molecule as a whole by using the formula
$E(R)=E_{\rm e} +1/R$.
These potential curves and corresponding values of $\Gamma/2$
are presented in Figure~\ref{foobar:HHPEFGRA}. 
One can
see in Figure~\ref{foobar:HHPEFGRA} that when growing electric field strength the
minimum on the curve $E(R)$ shifts to the right, becomes more shallow
and flat, and at $F_{\rm c}\approx0.065$ the minimum disappears. The
dependence of the location $R_0$ of the minimum on these curves with
corresponding values of the energy $E$ and
half-widths of the level are presented in Table~\ref{tabthree}. (For
comparison, at $F=0$ published results for the equilibrium point
are $R_0 =1.997193$ and $E=-0.6026346$ \cite{Bishop70}.
For the potential curve as a whole our numerical values at $F=0$
coincide with data by \cite{Sharp71}.)
One can see in Table~\ref{tabthree} and Figure~\ref{foobar:HHPEFGRA} 
that for the ground state of the hydrogen molecular ion at 
equilibrium internuclear distances there is a marked  
probability of its decay through the separation of the electron.  
Analogous calculations for two excited states $2p\pi_{\rm u}$ (the
lowest state with $|m|=1$) and $3d\delta_{\rm g}$ (the lowest state
with $|m|=2$) \cite{Ivanov98a} show that for the field strengths 
at which their potential curves have minima this 
process has a low probability.

\begin{table}
\caption{Equilibrium internuclear distances and corresponding energies
and half-widths of the energy of the ground state of
$\rm H_{2}^+$ molecule in
longitudinal electric fields. \label{tabthree}}
\footnotesize\rm
\begin{tabular}{@{}llllllllll}
$F$&$R_0$&$E_0$&$\Gamma /2$
&$F$&$R_0$&$E_0$&$\Gamma /2$
\cr
\noalign{\hrule}
0.00 &1.997&-0.60264&---      &0.04 &2.062&-0.60686&1.23(-19)\cr
0.01 &2.001&-0.60289&---      &0.05 &2.112&-0.60943&7.32(-15)\cr
0.02 &2.012&-0.60366&---      &0.06 &2.198&-0.61285&1.49(-11)\cr
0.03 &2.031&-0.60497&---      &0.065&2.28 &-0.61501&3.94(-10)\cr
\end{tabular}
\end{table}

The critical value of the maximal electric field
$F_{\rm c}=0.065{\rm a.u.}=3.3{\rm V/\AA}$ when the
molecule $\rm H_{2}^+$ can exist is in
a good agreement with experimental results by Bucksbaum {\it et al} 
\cite{Bucksbaum90}. 
According this work $\rm H_{2}^+$ molecule may exist in laser
beam fields with intensity less than $10^{14}{\rm W/cm^2}$ which
corresponds to $3{\rm V/\AA}$, and does not exist in more intensive
fields.

Other theoretical explanations of the rupture of $\rm H_{2}^+$ molecule in
intensive laser beam fields in works 
\cite{Bucksbaum90,Codling93,Yang91} were based on
conception of deformation of potential curves which results in
coupling of the ground state $1s{\sigma}_{\rm g}$ with
$2p{\sigma}_{\rm u}$ state \cite{Codling93}. This results
in a possibility of rupture of the molecule after absorption some
photons. On the other hand, the results obtained above enable us to
analyse this process from simpler point of view. When the frequency of
the radiation is low enough the consideration of the process as
extension and rupture of the molecule in a strong static electric
field is a quite adequate model \cite{Keldysh64,Codling93}.
The conception of coupling states might be substituted
by an exact numerical calculation of one (as it was done above) or
several (if several states are considered) dependencies $E(R)$. This
numerical calculation is equivalent to a traditional calculation of
the energy of the state when taking into account coupling with all the
states of the same symmetry (corresponding to the symmetry of the
initial physical problem).

\begin{figure}[b!] 
\centerline{\epsfig{file=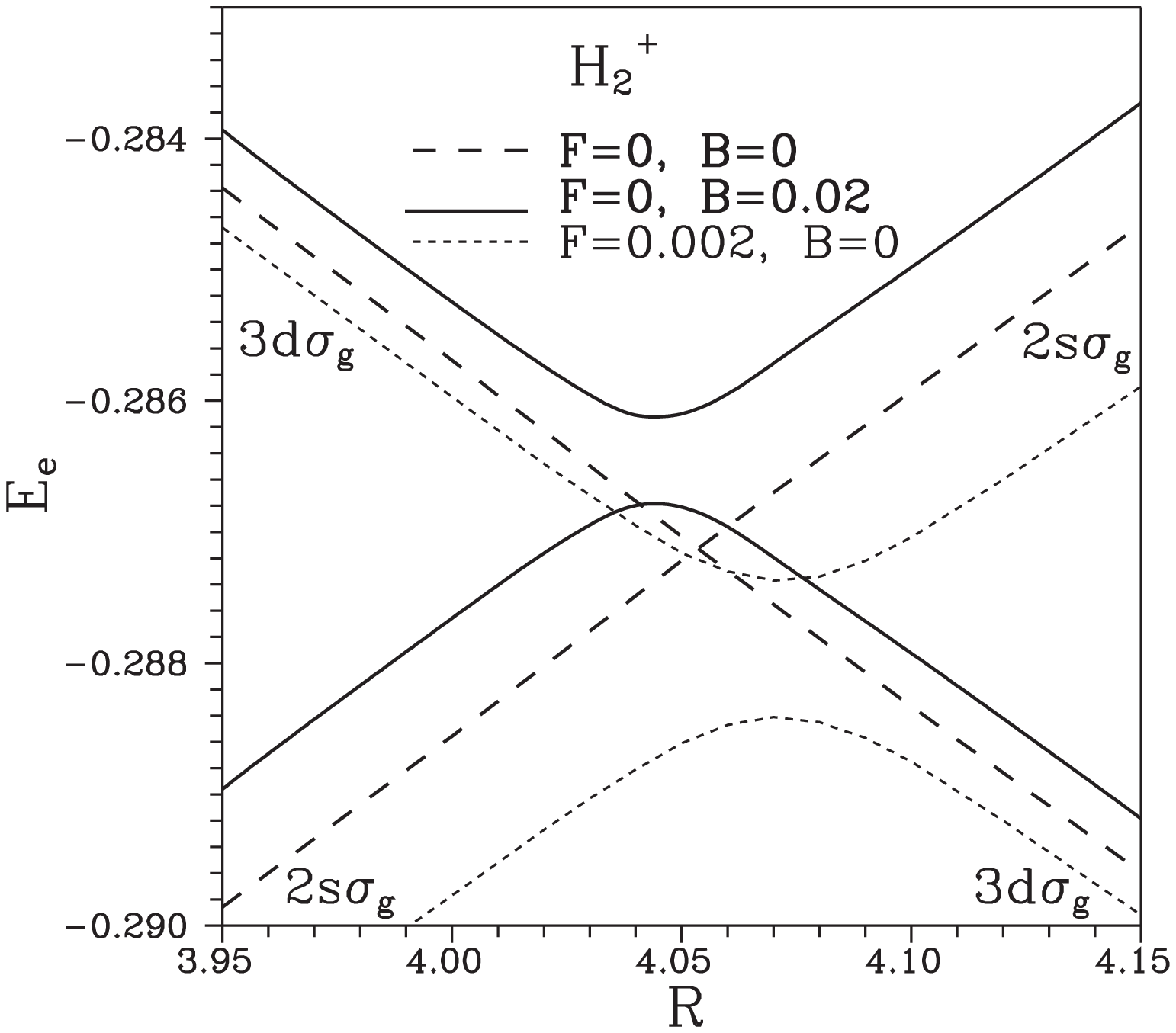,width=7cm}\hspace*{0.8cm}\epsfig{file=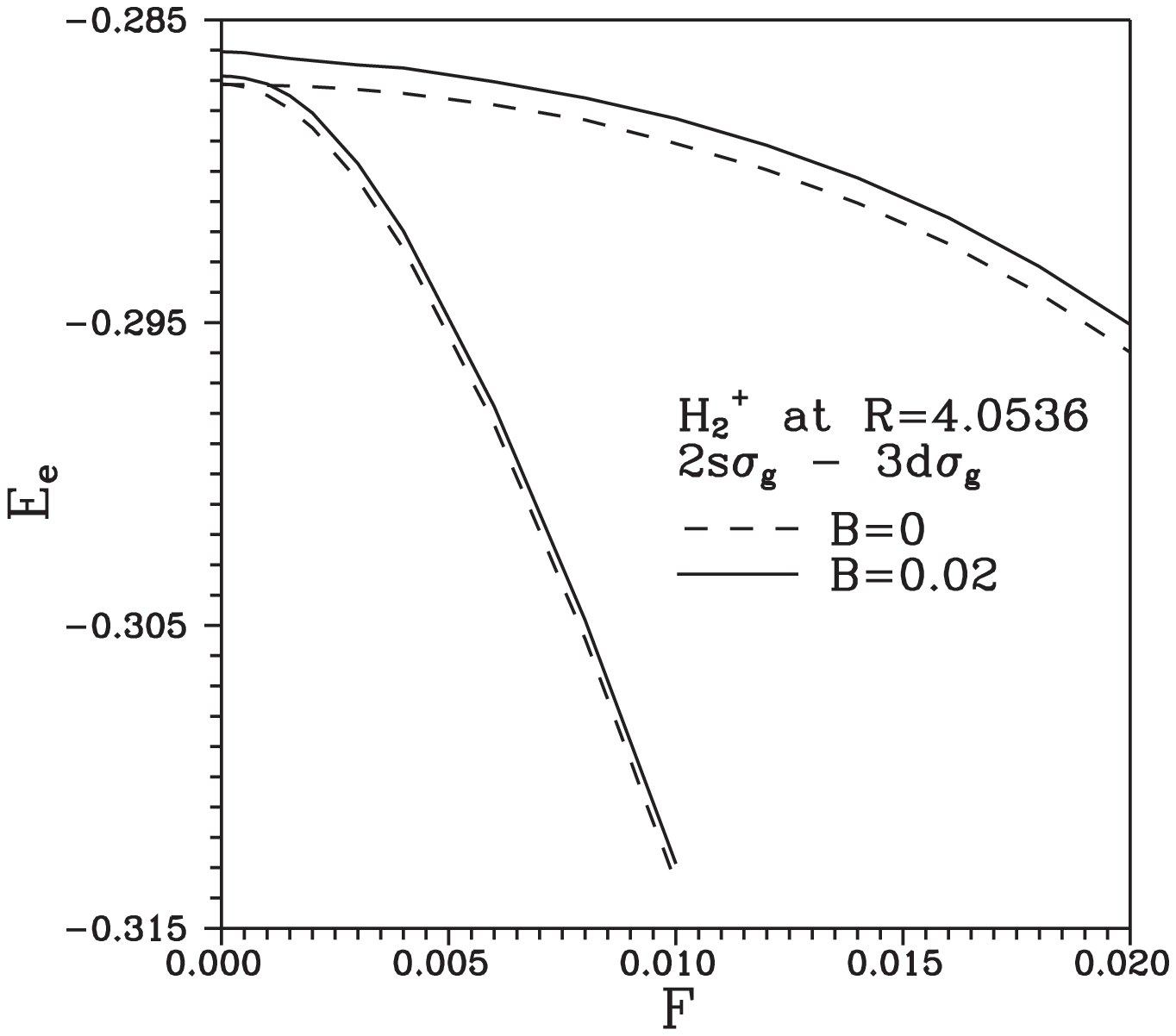,width=7cm}}
\vspace{10pt}
\caption{Left -- The electron energies $E_e$ of $2s\sigma_g$ and $3d\sigma_g$
states of the ${\rm H_2^+}$ molecular ion as functions of the
internuclear distance $R$ in the field-free space (broken),
in the magnetic field $ \gamma = 0.02 $ parallel to
the molecular axis (full) and in the electric field $F=0.002$
parallel to the molecular axis (dotted).
Right -- $E_e(F)$ for the doublet of $2s\sigma_g$ and $3d\sigma_g$
states of ${\rm H_2^+}$ at $R=4.0536$a.u. for $\gamma = 0 $
(broken) and $ \gamma = 0.02 $ (full).}
\label{foobar:H2FE}
\end{figure}

The calculations presented above were carried out
for the ground state of 2D Hamiltonian.
In fact, our method is not restricted by these states.
Computational algorithms of our program of atomic and molecular 
mesh calculations ATMOLMESH are constructed so that
calculations are being carried out for a state
with prescribed spatial symmetry
and with the electron energy nearest to the given initial
approximation $E_{\rm b}$.
Thus, there is no difference between calculations for
the ground and for excited states
and calculations can be fulfilled
for various states of the same
spatial symmetry including even degenerate states
(when applying some special technical methods).
For instance, potential curves
$2s\sigma_{\rm g}$ and $3d\sigma_{\rm g}$
of the field-free $\rm H_{2}^+$
have exact crossing near $R=4.05$ \cite{Sharp71}.
In magnetic fields parallel to the molecular axis this pair of states
forms an avoided crossing.
Our method permits calculations for both exact and
avoided crossings at $F=0$ as well as at $F \neq 0$
as presented in Figure~\ref{foobar:H2FE}.

The second singe-electron system which we present here is the
hydrogen atom in parallel electric and magnetic fields \cite{Ivanov2001}. 
Some numbers obtained for this system are shown in Table~\ref{tabHFB6}.

\begin{table}
\caption{The ground state of the
hydrogen atom in parallel electric and magnetic fields
at $F=0.1,\ 1,\ 5$.\label{tabHFB6}}
\footnotesize\rm
\begin{tabular}{@{}llllllllllll}
&\multicolumn{2}{c}{$F=0.1$}
&\multicolumn{2}{c}{$F=1$}
&\multicolumn{2}{c}{$F=5$}\cr
\cline{2-3}\cline{4-5}\cline{6-7}\\
$\gamma$&$E_0$&$\Gamma /2        $&$E_0$&$\Gamma /2         $&$E_0$&$\Gamma /2$\\
\noalign{\hrule}
$0   $&$-0.5274183$&$7.26904(-3)$&$-0.6243366$&$0.6468208 $&$-0.1350071$&$3.083929   $\\
$0.01$&$-0.532390 $&$7.2624(-3) $&$-0.629329 $&$0.646812  $&$-0.140005 $&$3.083925   $\\
$0.1 $&$-0.574600 $&$6.6392(-3) $&$-0.673584 $&$0.646053  $&$-0.184739 $&$3.083589   $\\
$1   $&$-0.8443098$&$9.5923(-5) $&$-1.0421379$&$0.577291  $&$-0.6077008$&$3.050207   $\\
$10  $&$-1.7498730$&---          &$-1.9579187$&$0.1173924 $&$-2.375552 $&$1.955678   $\\
$100 $&$-3.790110 $&---          &$-3.8219215$&$4.9988(-5)$&$-4.3806709$&$0.4909467  $\\
$1000$&$-7.66247  $&---          &$-7.66807  $&---         &$-7.82561  $&$1.04701(-2)$\\
\end{tabular}
\end{table}

\begin{figure}[b!] 
\centerline{\epsfig{file=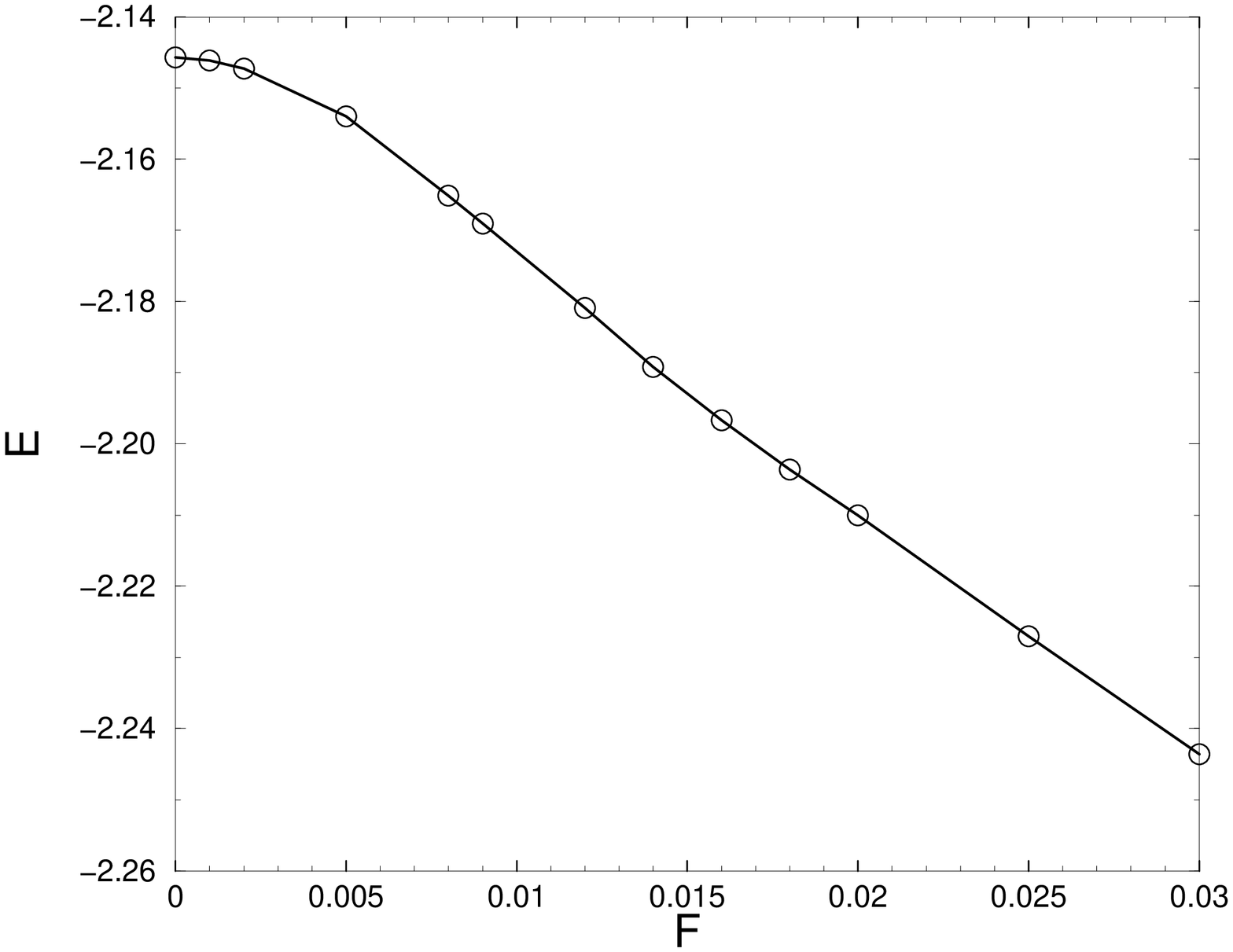,width=6cm,angle=270}
\hspace*{0.2cm}\epsfig{file=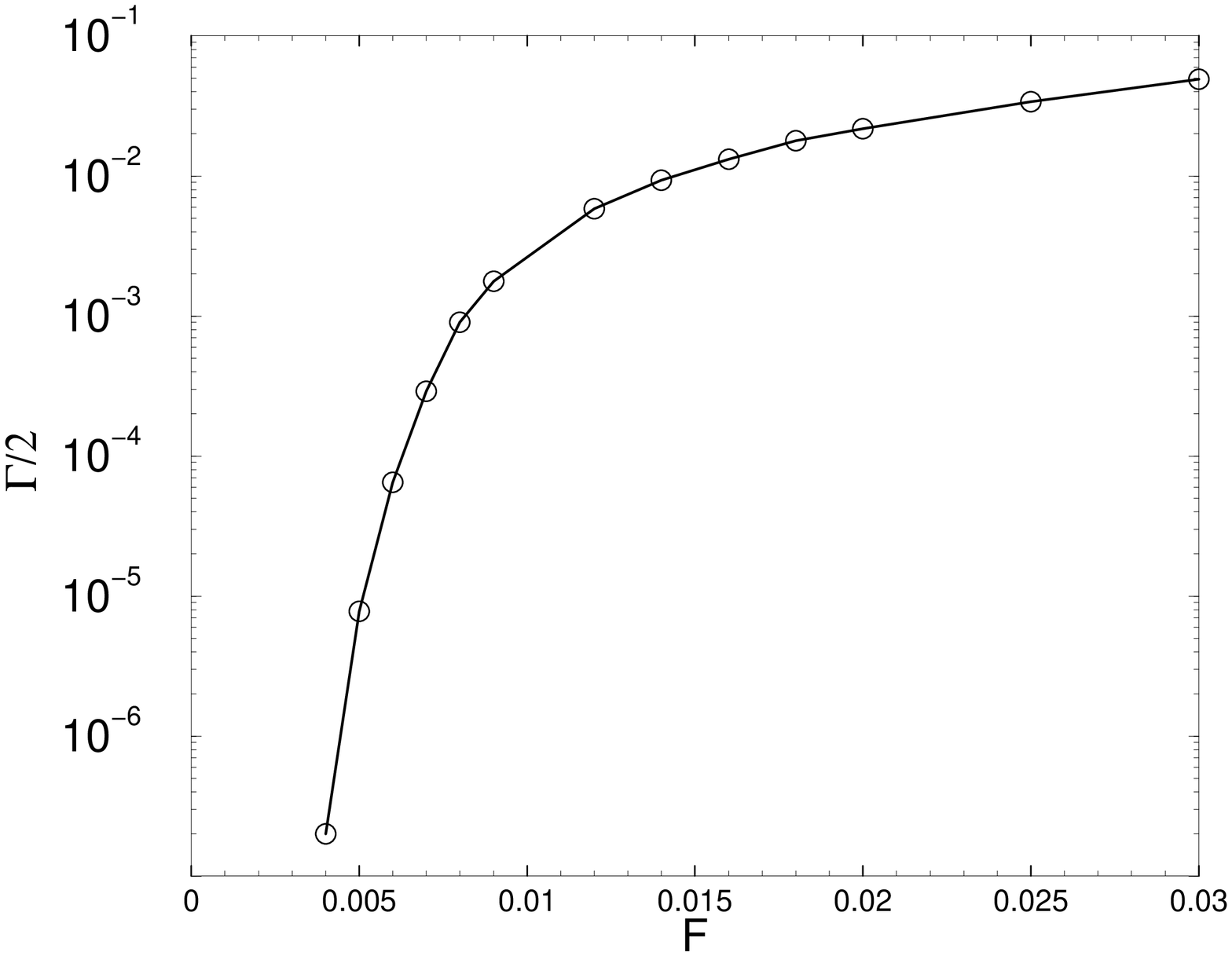,width=6cm,angle=270}}
\vspace{10pt}
\caption{Left -- The total energies of $2^1S_0$ 
state of the helium atom dependent on the electric field strength. 
Right -- Half-widths of this energy level.}
\label{foobar:He21SFE}
\end{figure}

Concluding this section we announce some preliminary results 
on the helium atom in strong electric fields. 
These results are obtained in a multi-configurational (CI) 
mesh calculation. 
The calculations for separate meshes are analogous to 
described above and are carried out by means of the
curved path exterior complex transformation. 
Our preliminary results for the ground state of the helium atom 
agree with data from recent works \cite{Scrinzi99,Themelis2000}. 
Our results for the excited state $2^1S_0$ are shown in 
Figure~\ref{foobar:He21SFE}.

\section{Conclusions}

In this communication we have presented a 2D fully numerical
mesh solution approach to calculations for atoms
and simple diatomic molecules in strong external electric
and magnetic fields.
For single-electron systems in external electric fields 
we can apply three different
methods of calculation of the complex energy eigenvalues
for atom-like systems and two methods for molecules.
These methods have different but overlapping regions
of their most effective application.
This allows their combined using for the reciprocal control
in applications.
The complex boundary condition method is the most straightforward and is
the most reliable in this sense.
For relatively weak fields it allows obtaining results
on the meshes with the lesser number of nodes than
the curved path complex transformation.
On the other hand, the boundary condition method loses its precision
and stability at extremely strong electric fields and, second,
this method contain the energy of the electron in its formulation.
The latter feature can be some obstacle in its application to
problems more complicated than the single-electron Schr\"odinger equation.
The curved path complex transformation method is the most
general of the three and the most prospective for
the following applications.
The only shortage of this method is that this is less precise and
less stable for the atom-like systems at extremely strong electric fields
($|{\rm Im} E | >> | {\rm Re} E|$) than the traditional complex rotation.
Thus, the latter method is the most convenient
for atom-like systems at such extremely strong fields.

The mathematical technique developed for solving
Schr\"odinger equations for quasi-steady states
allowed us to obtain a series of results for the hydrogen
atom in parallel electric and magnetic fields and
for the ${\rm H_2^+}$ ion in strong electric fields.
The following applications of these methods are associated with  
the CI approach which is now in the process of development and 
testing on the problem of the helium atom in strong electric fields.
We present preliminary results for the $2^1S_0$ state of this atom. 

{}


\begin{references}


\bibitem{Keldysh64}
Keldysh L. V., Sov. Phys. -- JETP, 47 (1964) 1945.


\bibitem{BenGre80}
Benassi L. and Grecchi V., J. Phys. B: At. Mol. Phys., 13
(1980) 911.

\bibitem{Franceschini}
Franceschini V., Grecchi V. and Silverstone H. J.,
Phys. Rev. A, 32 (1985) 1338.  

\bibitem{Froelich76}
Froelich P. and Br\"andas E., Int. J. Quantum Chem., S10 (1976) 1577.

\bibitem{Kolosov87}
Kolosov V. V., J. Phys. B: At. Mol. Phys., 20 (1987) 2359.

\bibitem{Kolosov89}
Kolosov V. V., J. Phys. B: At. Mol. Opt. Phys., 22 (1989) 833.


\bibitem{NicThem92}
Nicolaides C. A. and Themelis S. I., Phys. Rev. A, 45 (1992) 349.

\bibitem{NicGot92}
Nicolaides C. A. and Gotsis H. J., J. Phys. B: At. Mol. Opt. Phys., 25 (1992) L171.


\bibitem{AliHopf94}
Alijah A. and von Hopffgarten A., J. Phys. B: At. Mol. Opt. Phys., 27 (1994) 843.

\bibitem{Rao94}
Rao J., Liu W. and Li B., Phys. Rev. A, 50 (1994) 1916.

\bibitem{SilvermaN88}
Silverman J. N. and Nicolaides C. A., Chem. Phys. Lett., 153 (1988) 61.

\bibitem{Ivanov82}
Anokhin S. B. and Ivanov M. V., Sov. Phys. -- Solid
State, 24 (1982) 1979.

\bibitem{ZhVychMat}
Ivanov M. V., USSR Comput. Math. \& Math. Phys., 26 (1986) 140.

\bibitem{Ivanov88}
Ivanov M. V., J. Phys. B: At. Mol. Opt. Phys., 21 (1988) 447.

\bibitem{Ivanov91}
Ivanov M. V., Opt. Spektrosk., 70 (1991) 259;
English transl.: Opt. Spectrosc., 70 (1991) 148.

\bibitem{Ivanov94}
Ivanov M. V., J. Phys. B: At. Mol. Opt. Phys., 27 (1994) 4513.

\bibitem{Ivanov98}
Ivanov M. V., Phys. Lett. A, 239 (1998) 72.

\bibitem{IvaSchm98}
Ivanov M. V. and Schmelcher P., Phys. Rev. A, 57 (1998) 3793.

\bibitem{IvaSchm99}
Ivanov M. V. and Schmelcher P., Phys. Rev. A, 60 (1999) 3558.

\bibitem{IvaSchm2000}
Ivanov M. V. and Schmelcher P., Phys. Rev. A, 61 (2000) 022505.

\bibitem{Ivanov94a}
Ivanov M. V., Opt. Spektrosk., 76 (1994) 711;
English transl.: Opt. Spectrosc., 76 (1994) 631.

\bibitem{Ivanov98a}
Ivanov M. V., J. Phys. B: At. Mol. Opt. Phys., 31 (1998) 2833.

\bibitem{Ivanov2001}
Ivanov M. V. to be published.

\bibitem{Ivanov85}
Ivanov M. V., Sov. Phys. Semicond., 19 (1985) 1167.

\bibitem{Ivanov97}
Ivanov M. V., Opt. Spectrosc., 83 (1997) 23.

\bibitem{AbramSteg}
Abramowitz M. and Stegun I.A., Handbook of Mathematical Functions,
(Dover Publications, New York) 1972.

\bibitem{Ivanov83VLGU}
Anokhin S. B. and Ivanov M. V. Vestn. Leningr. Univ.
Fiz. \& Khim., no.2 (1983) 65.

\bibitem{Simon79}
Simon B., Phys. Lett. A, 71 (1979) 211.

\bibitem{Bishop70} 
Bishop D. M., J. Chem. Phys., 53 (1970) 1541.

\bibitem{Sharp71} 
Sharp T. E., Atomic Data, 2 (1971) 119.

\bibitem{Bucksbaum90}
Bucksbaum P. H., Zavriyev A., Muller H. G. and Schumacher D. W., Phys. Rev. Lett., 
64 (1990) 1883. 

\bibitem{Codling93}
Codling K. and Frasinski L. J., J. Phys. B: At. Mol. Opt. Phys., 26 (1993) 783.

\bibitem{Yang91}
Yang B., Saeed M., DiMauro L. F., Zavriyev A. and
Bucksbaum P. H., Phys. Rev. A, 44 (1991) R1458.

\bibitem{Scrinzi99}
Scrinzi A., Geissler M. and Brabec T. Phys. Rev. Lett., 83 (1999) 706.

\bibitem{Themelis2000}
Themelis S. I., Mercouris T. and Nicolaides C. A. Phys. Rev. A, 61 (2000) 024101.




\end{references}
\end{document}